\documentclass[aps,apl,twocolumn,showkeys,nofootinbib,reprint]{revtex4-1}

\usepackage{graphicx,amsmath,amssymb,amsfonts,latexsym,color,dcolumn,bm,mathtools,dsfont,siunitx}

\usepackage{pgfplots}
\usepackage{arydshln}
\usepackage[utf8]{inputenc} 
\usepackage{upgreek}
\usepackage{float}

\DeclareSIUnit\dBm{dBm}
\DeclareSIUnit\micron{\micro\metre}

\begin{document}


\title{Kinetic Inductance Traveling Wave Amplifiers For Multiplexed Qubit Readout}

\author{L. Ranzani$^1$,  M. Bal$^2$, Kin Chung Fong$^1$, G. Ribeill$^1$, X. 
Wu$^2$, 
J. Long$^2$, H.-S. Ku$^2$, R. P. Erickson$^2$, D. Pappas$^2$, and T. A. 
Ohki$^1$\footnote{Corresponding
author:
leonardo.ranzani@raytheon.com}}
\affiliation{$^1$Raytheon BBN Technologies, Cambridge, Massachusetts 02138, USA}
\affiliation{$^2$National Institute of Standards and Technology, Boulder, Colorado 80305, USA}
\date{\today}

\begin{abstract}

We describe a kinetic
inductance traveling-wave (KIT) amplifier suitable for superconducting quantum 
information measurements and characterize its wideband scattering and noise 
properties. We use  mechanical 
microwave switches to calibrate the four amplifier scattering parameters up to the device input and
output connectors  at the dilution refrigerator base temperature and a tunable
temperature load to characterize the amplifier noise. Finally, we demonstrate
the high fidelity simultaneous dispersive readout of two superconducting
transmon qubits. The KIT  amplifier provides low-noise amplification of both
readout tones with readout fidelities of 83\% and 89\% and negligible effect
on qubit lifetime and coherence.
%
\end{abstract}

\pacs{Valid PACS appear here}
\maketitle



Fast and high-fidelity readout of superconducting qubits is essential to
implementing complex quantum algorithms~\cite{riste2017demonstration}, error
correction~\cite{ofek2016extending,barends2014superconducting} and quantum
feedback~\cite{riste2013deterministic,vijay2012stabilizing}. The low-noise amplification of
readout signals is usually achieved via Josephson parametric amplifiers and
ring
modulators~\cite{yamamoto2008flux,castellanos2008amplification,bergeal2010phase,lecocq2017nonreciprocal};
 however
lumped-element  Josephson parametric amplifiers (JPAs) typically have low saturation power and bandwidths
of a few megahertz that can be extended up to 700~\si{MHz} with suitable impedance matching 
techniques~\cite{mutus2014strong,eichler2014quantum,roy2015broadband,simoen2015characterization}.
 As the size and complexity of superconducting
quantum circuits
increases~\cite{chen2014qubit,riste2017demonstration,takita2017experimental}, it
is desirable to extend the amplifier bandwidth and saturation power in order  to 
achieve higher measurement speeds by performing the simultaneous
dispersive readout of a large number of qubits, while simultaneously increasing the 
power of each individual readout tone. Traveling-wave parametric amplifiers 
provide such an option because of their wide bandwidth, typically spanning 
several gigahertz, high saturation power and near quantum limited
noise performance. In this case, amplification is obtained by injecting the input signal
together with a strong co-propagating pump into a nonlinear transmission
medium, consisting of either a long array of Josephson
junctions or 
SQUIDs~\cite{macklin2015near,o2014resonant,bell2015traveling,white2015traveling,zorin2016josephson,zorin2017traveling},
 or of a high kinetic inductance 
material~\cite{eom2012wideband,vissers2016low,erickson2017theory,chaudhuri2017broadband}.
In both cases the propagating pump modulates the inductance per unit length of
the line and parametrically amplifies the weak input signals. Kinetic inductance 
traveling-wave (KIT) amplifiers
further provide an extremely high saturation power ($>\SI{-60}{\dBm}$),  in 
addition to the
broad ($\sim$\SI{4}{GHz}) bandwidth, that make them a desirable tool for the
simultaneous readout of superconducting qubits. Moreover, recent demonstrations 
of the KIT amplifier have significantly lowered the required pump power, thus 
enabling operation near sensitive quantum devices~\cite{vissers2016low}. \\

In this work we characterize the gain, return loss and noise temperature of a KIT
amplifier at 10~mK optimized for superconducting qubit readout. The amplifier has 
relatively
low pump power \SI{-30}{\dBm}, \SI{12}{\decibel} gain with low ripple and wide bandwidth. Moreover we obtain an
in-band gain ripple lower than \SI{3}{\decibel} between 4 and \SI{12}{GHz} and system noise
temperature as  low as \SI{1.5}{K}. This is about a factor of ten lower than the system noise of a typical superconducting qubit 
measurement chain with HEMT amplifiers only, which is 10-20~K. We further demonstrate the simultaneous readout of
two  superconducting transmon qubits with readout fidelities up to 89\%
and  show that the amplifier has no effect on qubit lifetime and
coherence within our measurement errors. Furthermore the large saturation power~\cite{vissers2016low}
(-40~dBm) of the KIT amplifier makes it a suitable candidate for the multiplexing of superconducting qubit readout tones.\\
%

For our measurements we operated a KIT amplifier in three-wave mixing
mode~\cite{vissers2016low,erickson2017theory}. The device consists of a meandered 
superconducting
coplanar waveguide lithographically defined on a silicon chip. The 
superconductor exhibits a high kinetic inductance that depends nonlinearly on the 
current: 
\begin{equation}
L_k(I)\approx L_k(0)\left[1+\left(\frac{I}{I_*}\right)^2\right],
\end{equation}
where the current $I_*$ is comparable to the critical current $I_c$ of the
superconductor\cite{eom2012wideband}. By injecting a strong microwave pump at 
frequency $\omega_p$
into the line we induce a radio-frequency modulation of the current
$I_{rf}=a_p\cos(k_px-\omega_pt+\phi)$ and, therefore, of the line  inductance
$L_k(x,t)$. In three-wave mixing operation we further introduce a bias current
$I=I_{rf}+I_{dc}$, so that the signal and idler mode frequencies satisfy the 
constraint $\omega_s+\omega_i=\omega_p$. The
amplifier gain grows exponentially with the device length under the phase
matching condition $k_s+k_i=k_p-\Delta \theta$, where $\Delta
\theta$ is the nonlinear phase shift of the pump due to self- and cross-phase 
modulation.
For $I_{dc} \ll I_*$, the gain is $G\approx 1+\sinh^2(gL)$,
with~\cite{eom2012wideband,erickson2017theory}:
\begin{equation}
g=\sqrt{k_sk_i}\frac{I_{dc}}{2I_*^2}a_p,
\end{equation}
In a uniform coplanar waveguide (CPW) line in general $k_p \approx k_s+k_i$ and 
the phase matching
condition is not satisfied unless artificial dispersion is introduced by 
periodically
modulating the CPW line width to introduce a bandgap near the pump
frequency. Finally we observe that, even when operating the KIT amplifier under a 
nonzero dc current bias, 4-wave mixing and other parametric processes are present 
and impact the device 
characteristics~\cite{erickson2017theory}.  
\begin{figure}
		\includegraphics[width=8.5cm]{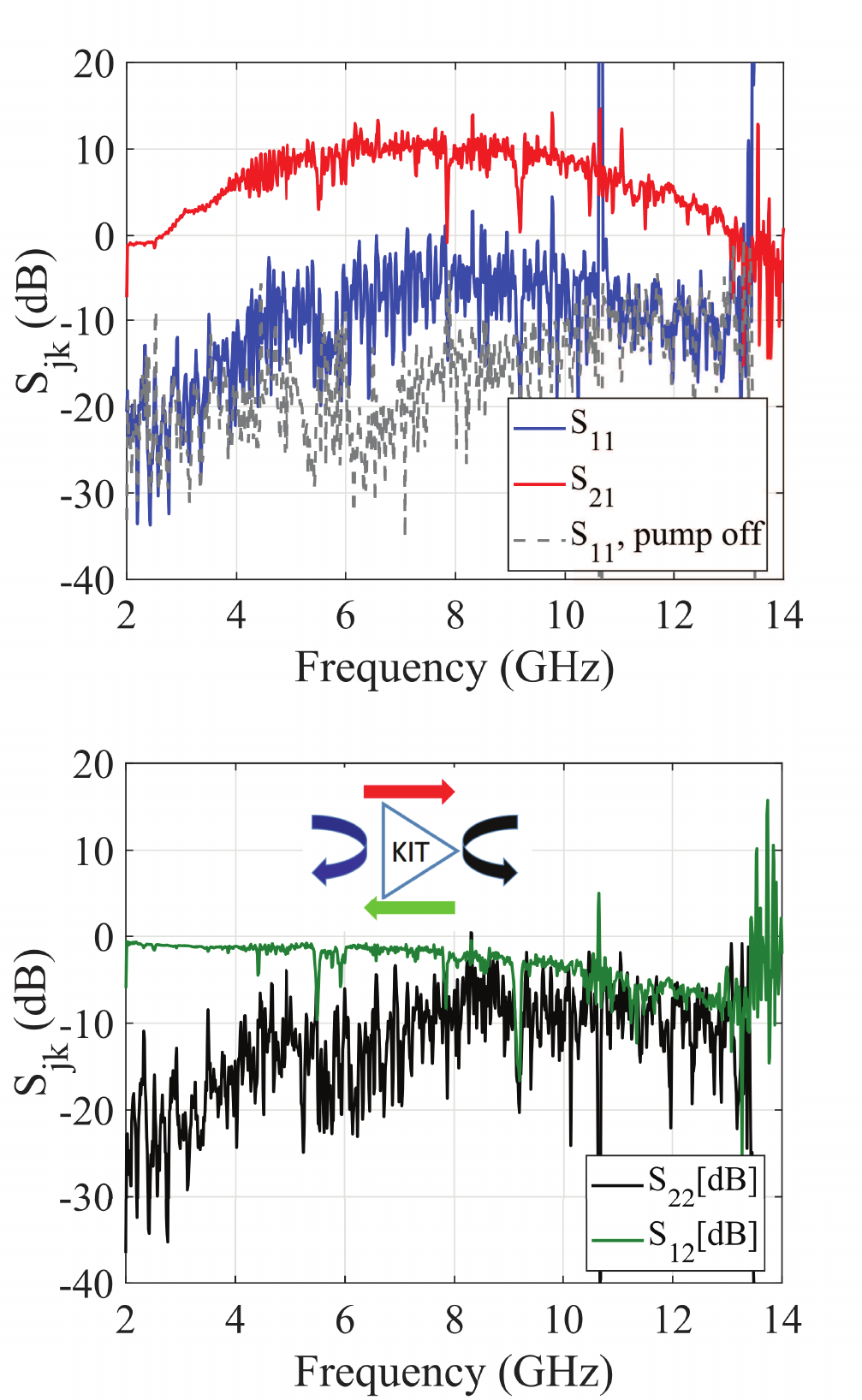}
\caption{KIT scattering parameters: forward gain and input return loss (left) and 
reverse gain and output return loss (right). The return loss increases when the 
gain is turned on, because the device reverse transmission and gain create a feedback mechanism that amplifies the reflected 
signals.
	\label{fig1}}
\end{figure}
Our KIT amplifier is fabricated on a 20nm NbTiN film over a
2.0$\times$2.2~\si{cm}, \SI{381}{\micron} thick  silicon chip. The meandered CPW 
line is \SI{2}{m} long, has a gap of \SI{2}{\micron} and width of 
\SI{3}{\micron}, which is increased to 
\SI{6}{\micron} at the
periodic loadings used for dispersion engineering. Since the line characteristic
impedance is \SI{180}{\ohm}, we use 50 to 180~\si{\ohm} triangular tapers 
for impedance matching. Furthermore, we coat the ground plane with gold to 
suppress parasitic microwave modes and thermalize the device. 

We characterize the amplifier gain and return loss in our dilution
refrigerator at a temperature T$\sim$\SI{10}{mK}. The KIT amplifier is
connected on both sides to single-layer Nb superconducting microwave diplexers to 
separate the signal (4-12~\si{GHz})
from the parametric pump (17~\si{GHz}), while the bias current $I_{dc}$ is 
provided via
commercial external bias tees. The superconducting diplexers are fabricated in a 
single Niobium layer on a 6$\times$8.5~\si{mm} silicon 
chip. They consist of a low-pass stepped-impedance microstrip filter 
(dc-12~\si{GHz}, with $<$0.2~\si{\decibel} insertion loss up to 9~\si{GHz}) and a 
band-pass parallel coupled-lines (hairpin) filter 
(14-20~\si{GHz} with 0.4~\si{\decibel} of insertion loss at the pump 
frequency)~\cite{hunter2001theory}. The lowpass filter suppresses the pump power 
by more than 50~\si{\decibel}.

We use a set of cold thru-reflect-line (TRL) coaxial
calibration standards to deembed the intervening components in the measurement
chain and obtain the device scattering parameters~\cite{ranzani2013two}. A pair 
of 6-port mechanical dc-18~\si{GHz}
switches mounted onto the mixing chamber plate allows the selection of different
calibration standards without needing to warm up the system. We connect
the KIT and calibration standards to the switches via short
($\sim$6~\si{in}) identical coaxial cables to minimize calibration errors due 
to variations in the standard measurement lines. We then measure 
all four scattering parameters via two independent attenuated input lines
(\SI{90}{\decibel} loss) and two amplified output lines. By 
measuring three calibration standards, we extract
an error model for the intervening components before and after the
device under test and de-embed the scattering 
parameters of the KIT amplifier: the error model consists of 8 independent terms 
determined by a constrained interior-point nonlinear optimization 
algorithm~\cite{ranzani2013two}.\\

When the parametric pump is turned off and the KIT is biased with a bias current of
$I_b=1.5mA$, we measure the insertion loss of the CPW line and obtain
\SI{0.5}{\decibel} at \SI{2}{GHz} up to \SI{3}{\decibel} at \SI{8}{GHz}. We
then turn the  pump on and obtain the scattering parameters shown in
Figure~\ref{fig1}. We measure a peak forward gain of \SI{12}{dB} at
\SI{7.6}{GHz} with a \SI{3}{\decibel} bandwidth of \SI{8}{GHz} and $\pm$\SI{1.5}{\decibel} ripple. The reverse
gain of the device is independent on the pump power and equal to the device
insertion loss measured with no parametric pump, as expected. In
Figure~\ref{fig1} we also show the device return loss, which, interestingly, is
a function of the pump power. With no microwave pump applied to the KIT, we
measure a return loss of around \SI{20}{\decibel} on both ports, with a maximum
return loss of \SI{10}{\decibel} above \SI{8}{GHz}. However, as soon as the
microwave pump is turned on, the device reflection coefficients increase up to
$\sim \SI{0}{\decibel}$ around \SI{8}{GHz}, as shown in Figure~\ref{fig1}. We
can understand this behavior by observing that the amplifier impedance mismatch
is due to the interference of multiple reflections at the taper junctions as
well as at the multiple bends of the CPW line. When we tune the amplifier up,
the signal is amplified before each reflection, causing an increase in the total
return loss. If we further increase the pump power we obtain increase in gain ripple. Therefore, the low reverse 
isolation ($>\SI{-2.5}{\decibel}$) and
impedance match effectively limit the maximum amount of gain that
can be extracted from the device. Nevertheless, we show below that our KIT 
amplifier still provides a substantial improvement in qubit readout
fidelity.\\
\begin{figure}
	\includegraphics[width=7.5cm]{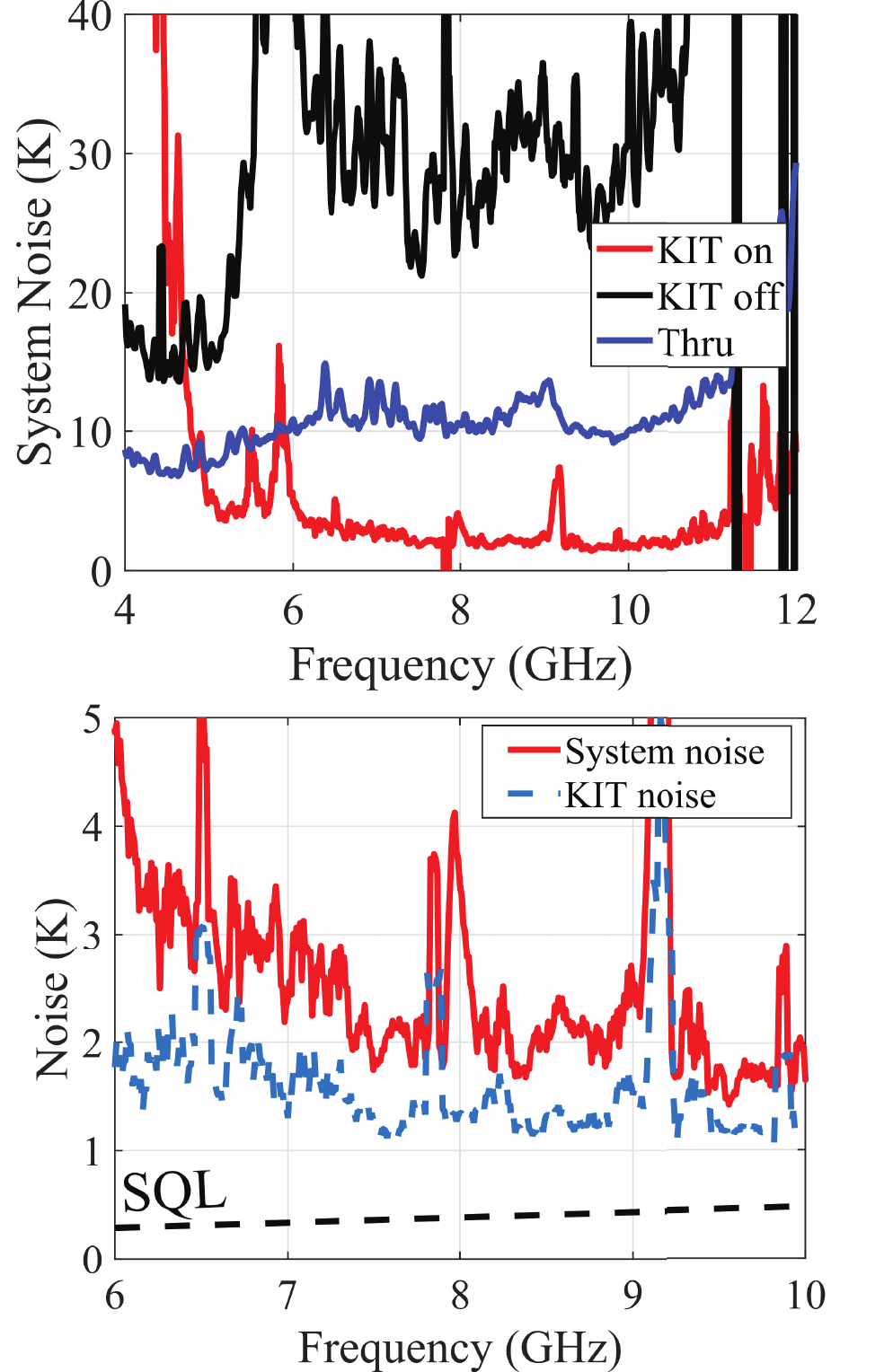}
	\caption{Amplifier noise temperature measurement acquired with a variable 
	temperature load tuned between \SI{300}{mK} and \SI{3}{K}. The system noise 
	temperature without the amplifier (blue line) is compared to the noise 
	temperature measured with the amplifier turned off (black line) and on (red 
	line). The KIT amplifier provides a factor of 10 improvement in system noise. 
	The figure on the right shows a detail of the system noise and the KIT noise 
	after removing the effect of the external 
	components (blue line).  
		\label{fig2}}
\end{figure}
\begin{figure}
	\includegraphics[width=8cm]{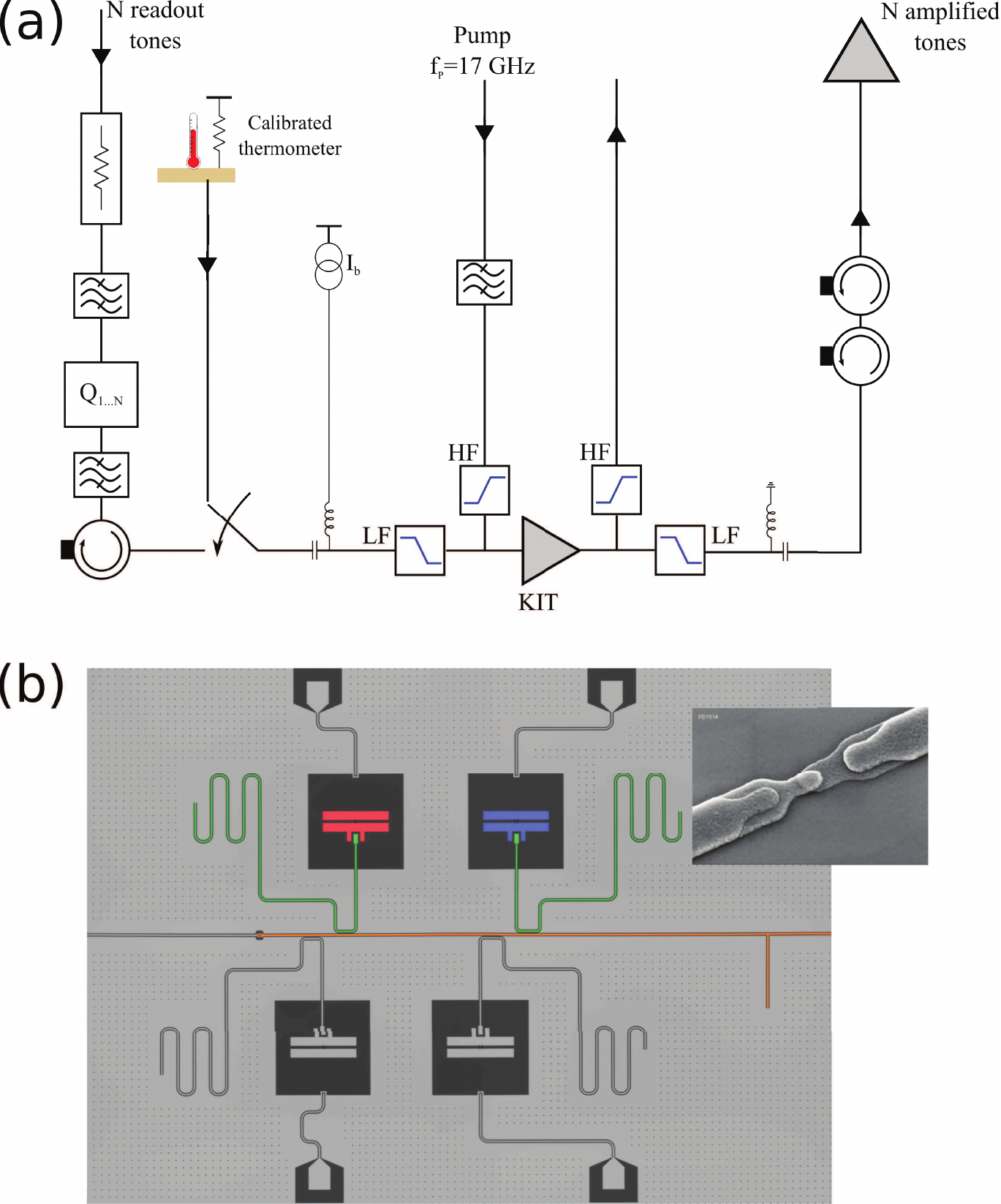}
	\caption{(a) Qubit readout setup: the qubit chip is measured in transmission
		with N readout tones that are then directed to a KIT amplifier, anchored
		to
		the mixing chamber plate, followed by a two-stage circulator and a HEMT
		amplifier. (b) Layout of the qubit chip consisting of 4 transmon qubits,
		quarter-wave readout resonators (green), and readout Purcell filter
		(orange). We measured qubit
		$Q_2$ (red) and $Q_3$ (blue). The readout resonators are placed in the 
		center of the Purcell
		filter
		band (7.1-7.4~GHz), while the transmon qubits are placed around 5.5~GHz.}
	\label{fig3}
\end{figure}
To characterize the noise temperature, we perform
a hot/cold load measurement by use of a variable temperature load
noise source, which is heated by injecting a dc
current and measured by a thermocouple. The source is anchored to the cold plate 
via a low thermal conductivity platform. The
temperature of the noise source can be continuously tuned from \SI{100}{mK} to
\SI{3}{K}, unlike a traditional hot/cold measurement where only two temperature
points are available, thus leading to improved accuracy. We measure
the output noise spectral density $S(\omega,T)$ of the amplifier as a function of 
frequency $\omega$ and temperature $T$ in a
spectrum analyzer and compute the system noise via a minimum least
squares fit to:
\begin{equation}
S(\omega,T)=G\left(\frac{(\hbar\omega/k_B)}{e^{\hbar\omega/k_BT}-1}+T_{sys}(\omega)\right)
\end{equation}  
Note that the system noise
is characterized over the entire amplifier band in a single measurement, which is 
important for devices such as the KIT that amplify over several GHz
of bandwidth. In Figure~\ref{fig2} we show the measured system noise in Kelvin
as a function of frequency: we measure
$T^{off}_{sys}$=6-15~\si{K} between 4-12~\si{GHz} when the KIT is removed 
(blue line) and a lower system noise $T^{on}_{sys}$= 3-1.7~K when the KIT is 
inserted. The noise is referred to
the input of the amplifier chain. We also observe a noise
temperature increase when the amplifier is off (black line), due to the
insertion loss of the diplexers and bias tees (\SI{3}{\decibel} at \SI{8}{GHz}) 
and the amplifier itself (\SI{1}{\decibel} at \SI{8}{GHz}). In future device 
implementations these components could be
placed into the same package to minimize insertion loss and improve
impedance match. At \SI{9}{GHz}, where the gain is maximum, the system noise
$T^{on}_{sys}$=\SI{1.5}{K}($\pm$0.17K) corresponds to 3.5 noise photons. We 
attribute the excess noise to losses in the measurement chain, particularly in the
diplexers and bias tees, HEMT amplifier noise and thermal
heating of the KIT amplifier.  We use a distributed 
model~\cite{macklin2015near} of the KIT
internal gain and loss to estimate the intrinsic amplifier input noise 
and obtain 1.4 photons of added noise at \SI{10}{GHz} (close to the standard 
quantum limit)
and 4.1 photons at \SI{6}{GHz}.
Finally, we test the performance of the KIT amplifier in a single and two-qubit dispersive
readout experiment. In our measurement setup, shown in Figure~\ref{fig3}(a), we
replace the noise source in the previous experiment with a qubit circuit. This
circuit (Figure~\ref{fig3}(b)) consists of four transmon qubits
dispersively coupled to quarter-wave CPW readout resonators (center frequencies
7.1-7.4~\si{GHz}, coupling strength $g = \SI{38}{MHz}$). The readout resonators
are in turn capacitively coupled to a quarter-wave resonator Purcell
filter~\cite{jeffrey2014fast}, with readout resonator
damping rate $\kappa^{-1}=\SI{33}{ns}$ to achieve fast readout while preserving qubit
coherence. The filter is weakly coupled to an input
line via a small (15~fF) coupling capacitor and strongly inductively coupled
(external quality factor Q=22) to an output line to enable measurements in
transmission. We measured fixed frequency qubit $Q_2$ with 0-1 transition frequency
\SI{4.518}{GHz} and tunable $Q_3$ with 0-1 transition frequency \SI{5.772}{GHz}. Both
qubits have an anharmonicity of \SI{-310}{MHz}. Typical qubit coherence times
measured for this device are $T_1 = \SI{12}{\micro\second}$, $T_2 =
\SI{14}{\micro\second}$ and Hahn echo time $T_{\mathrm{echo}} =
\SI{19}{\micro\second}$.

\begin{figure}
	\includegraphics[width=8cm]{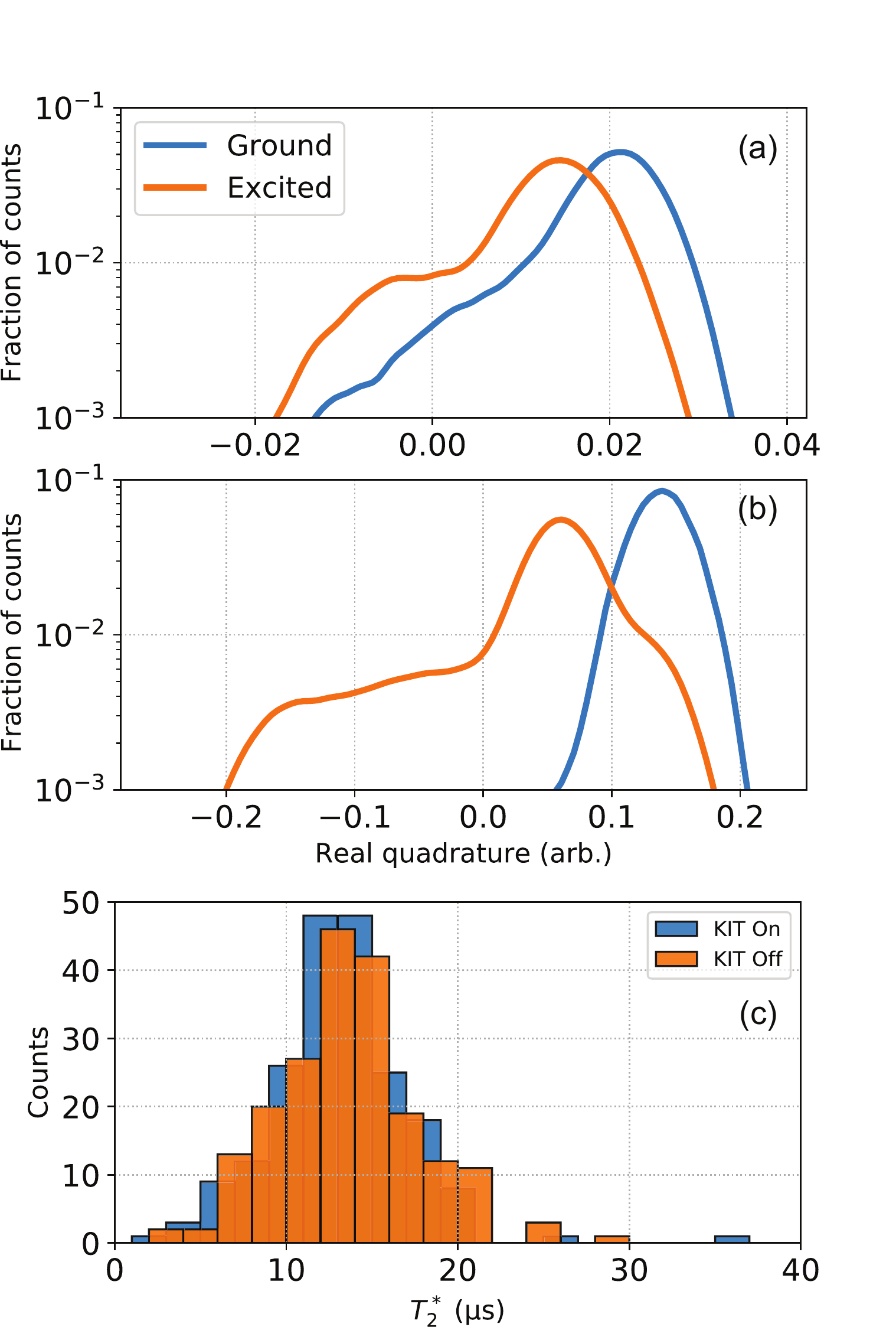}
	\caption{Qubit measurement data for qubit $Q_2$. (a) Histogram of measurement results with the KIT amplifier turned off, corresponding to $F = 71.7\%$. (b) Histogram of measurement results with the KIT amplifier turned on, corresponding to $F = 90.3\%$. (c) Histogram of $T_2^*$ measured with Ramsey interferometry on qubit $Q_2$, with the KIT on and off.
		\label{fig4}}
\end{figure}

In a first experiment we measure coherence and dephasing times $T_1$ and $T_2$
of qubit $Q_2$ with the KIT pump turned on and off for identical readout
drive parameters. We measure a slight decrease in
$T_1$ when the KIT is turned on, and no change in $T_2$ ($T_2^{off}$=13.6 $\pm$ 0.57 $\mu s$, $T_2^{on}$=12.85 
$\pm$ 0.42 $\mu s$, see Figure \ref{fig4}(c)). We fit to two frequency components 
in the Ramsey fringe experiments
used to measure $T_2$ and we ascribe a small splitting $\Delta f \sim \SI{600}{kHz}$,
consistent with charge dispersion of a transmons with $E_J/E_C = 30$ to $\pm e$
offsets induced by background fluctuations. \cite{schreier2008chargenoise,
riste2013fluctuations}. We note that the distribution of $T_2^*$ measured through
Ramsey interferometry is not affected by turning the KIT on, see Figure 
\ref{fig4}c, and verify that the Hahn echo time is also preserved 
($T_2^{e,off}$=18.52 $\pm$ 1.27 $\mu s$, $T_2^{e,on}$=21.19 
$\pm$ 3.38 $\mu s$). Finally, we do not observe a measurable Stark shift induced 
by the amplifier pump leaking into the readout resonator, consistent with 
the pump being far from the readout resonator 
center frequency and pump leakage being further suppressed by the diplexer, 
isolator and Purcell filter.

In a second experiment we perform single shot readout measurements of the same
qubit to determine readout fidelity. We prepare the qubit in its $|0\rangle$ or
$|1\rangle$ states \num{3e4} times, and monitor the transmission across the
Purcell filter while driving the qubit cavity on the dressed $|0\rangle$ with a
\SI{1}{\micro\second} long readout pulse. We use an optimal matched filter
approach \cite{ryan2015tomo} to integrate the digitized heterodyne signal and
rotate qubit state information into the real quadrature of the signal. Binning
this data, we measure well separated histograms corresponding to the two state
preparations (Figure~\ref{fig4}). We then extract the readout fidelity,
defined as $F=1-(P_{1|0|}+P_{0|1})/2$, where $P_{0|1}$ is the probability of
erroneously identifying the qubit state as $|0\rangle$ instead of $|1\rangle$, by
integrating the two histograms and taking the difference. We obtain a
fidelity of $F$=71.7\% with the KIT turned off, compared to 90.3\%
with the KIT turned on, corresponding to a 18.6\% improvement.
\begin{figure}
	\includegraphics[width=8cm]{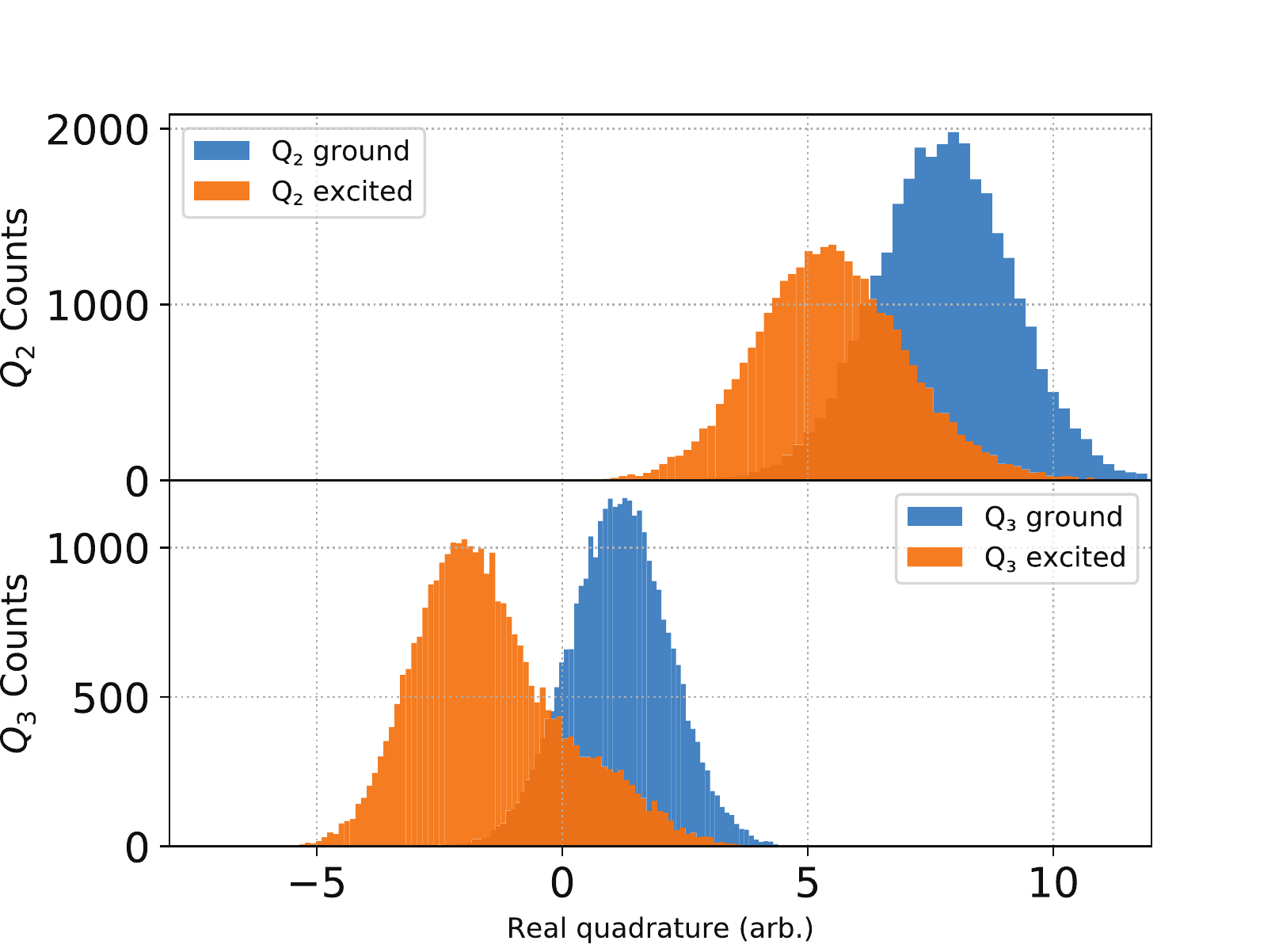}
	\caption{Two qubit fidelity histograms for 30,000 ground and excited state 
	preparations for $Q_2$ (top) and $Q_3$ (bottom). 
		Measurement fidelity is $F = 89.4\%$  for $Q_2$ and $F = 83.3\%$ for 
		$Q_3$. The individual error probabilities are 
		$P_{1|0|}$=7.01\% and $P_{0|1|}$=13.6\% for $Q_2$ and $P_{1|0|}$=14.5\% 
		and $P_{0|1|}$=17.8\% for $Q_3$. 
		\label{fig5}}
\end{figure}

Finally we perform a simultaneous readout of two qubits $Q_{2,3}$ by probing their
respective readout cavities at the same time. From the measured
histograms (Figure~\ref{fig5}) we extract readout fidelities of
89.4\% and 83.3\%. While these fidelities do not match the best results
achieved for other quantum-limited amplifiers~\cite{heinsoo2018rapid,gu2017microwave}, we believe that they demonstrate
the KIT promise for superconducting qubit readout, particularly for those
situations that benefit from its broad bandwidth and high saturation power. The 
fidelities reported
here are limited by the low gain of present KIT devices and the use of a qubit 
device not optimized for 
high-fidelity readout
($\kappa \gg 2\chi \approx \SI{2}{\mega\hertz}$).
In conclusion we characterized the gain and noise temperature of a KIT
amplifier at \SI{10}{mK}, demonstrating 12 $\pm$ \SI{1.5}{\decibel} gain and
\SI{1.5}{K} of system noise at \SI{9}{GHz}. Furthermore, we demonstrated the 
single shot
readout of superconducting qubits with a KIT amplifier and measured more than
90\% fidelity (single qubit) and 83-89\% (two-qubit). 
Further improvements need higher amplifier gain to further overcome 
microwave losses and HEMT noise. The maximum gain is limited by the 
insurgence of instability as the pump power is increased past a critical threshold. Higher gains 
require a further investigation into the effect of impedance fluctuations in the 
superconducting line, increasing the 
nonlinearity at low pump powers and use 50~$\Omega$ artificial 
lines~\cite{adamyan2016superconducting} that achieve better
impedance match by reducing overall device size and eliminating impedance 
transformers.

\section{Acknowledgments}
We acknowledge funding to BBN by NASA JPL (subcontract \#1548616) and to NIST by 
the Quantum Based Metrology Initiative (NIST), the Intelligence Advanced Research 
Projects Agency (IARPA) LogiQ Program, and  the ARO quantum computing program. 
R.P.E. acknowledges grants 60NANB14D024 and 70NANB17H033 from the US Department 
of Commerce. This work is a contribution of the 
U.S. Government, not subject to copyright.

\begin{thebibliography}{34}%
	\makeatletter
	\providecommand \@ifxundefined [1]{%
		\@ifx{#1\undefined}
	}%
	\providecommand \@ifnum [1]{%
		\ifnum #1\expandafter \@firstoftwo
		\else \expandafter \@secondoftwo
		\fi
	}%
	\providecommand \@ifx [1]{%
		\ifx #1\expandafter \@firstoftwo
		\else \expandafter \@secondoftwo
		\fi
	}%
	\providecommand \natexlab [1]{#1}%
	\providecommand \enquote  [1]{``#1''}%
	\providecommand \bibnamefont  [1]{#1}%
	\providecommand \bibfnamefont [1]{#1}%
	\providecommand \citenamefont [1]{#1}%
	\providecommand \href@noop [0]{\@secondoftwo}%
	\providecommand \href [0]{\begingroup \@sanitize@url \@href}%
	\providecommand \@href[1]{\@@startlink{#1}\@@href}%
	\providecommand \@@href[1]{\endgroup#1\@@endlink}%
	\providecommand \@sanitize@url [0]{\catcode `\\12\catcode `\$12\catcode
		`\&12\catcode `\#12\catcode `\^12\catcode `\_12\catcode `\%12\relax}%
	\providecommand \@@startlink[1]{}%
	\providecommand \@@endlink[0]{}%
	\providecommand \url  [0]{\begingroup\@sanitize@url \@url }%
	\providecommand \@url [1]{\endgroup\@href {#1}{\urlprefix }}%
	\providecommand \urlprefix  [0]{URL }%
	\providecommand \Eprint [0]{\href }%
	\providecommand \doibase [0]{http://dx.doi.org/}%
	\providecommand \selectlanguage [0]{\@gobble}%
	\providecommand \bibinfo  [0]{\@secondoftwo}%
	\providecommand \bibfield  [0]{\@secondoftwo}%
	\providecommand \translation [1]{[#1]}%
	\providecommand \BibitemOpen [0]{}%
	\providecommand \bibitemStop [0]{}%
	\providecommand \bibitemNoStop [0]{.\EOS\space}%
	\providecommand \EOS [0]{\spacefactor3000\relax}%
	\providecommand \BibitemShut  [1]{\csname bibitem#1\endcsname}%
	\let\auto@bib@innerbib\@empty
	\bibitem [{\citenamefont {Rist{\`e}}\ \emph {et~al.}(2017)\citenamefont
		{Rist{\`e}}, \citenamefont {Da~Silva}, \citenamefont {Ryan}, \citenamefont
		{Cross}, \citenamefont {C{\'o}rcoles}, \citenamefont {Smolin}, 
		\citenamefont
		{Gambetta}, \citenamefont {Chow},\ and\ \citenamefont
		{Johnson}}]{riste2017demonstration}%
	\BibitemOpen
	\bibfield  {author} {\bibinfo {author} {\bibfnamefont {D.}~\bibnamefont
			{Rist{\`e}}}, \bibinfo {author} {\bibfnamefont {M.~P.}\ \bibnamefont
			{Da~Silva}}, \bibinfo {author} {\bibfnamefont {C.~A.}\ \bibnamefont 
			{Ryan}},
		\bibinfo {author} {\bibfnamefont {A.~W.}\ \bibnamefont {Cross}}, \bibinfo
		{author} {\bibfnamefont {A.~D.}\ \bibnamefont {C{\'o}rcoles}}, \bibinfo
		{author} {\bibfnamefont {J.~A.}\ \bibnamefont {Smolin}}, \bibinfo {author}
		{\bibfnamefont {J.~M.}\ \bibnamefont {Gambetta}}, \bibinfo {author}
		{\bibfnamefont {J.~M.}\ \bibnamefont {Chow}}, \ and\ \bibinfo {author}
		{\bibfnamefont {B.~R.}\ \bibnamefont {Johnson}},\ }\href@noop {} 
		{\bibfield
		{journal} {\bibinfo  {journal} {npj Quantum Information}\ }\textbf 
		{\bibinfo
			{volume} {3}},\ \bibinfo {pages} {16} (\bibinfo {year} 
			{2017})}\BibitemShut
	{NoStop}%
	\bibitem [{\citenamefont {Ofek}\ \emph {et~al.}(2016)\citenamefont {Ofek},
		\citenamefont {Petrenko}, \citenamefont {Heeres}, \citenamefont 
		{Reinhold},
		\citenamefont {Leghtas}, \citenamefont {Vlastakis}, \citenamefont {Liu},
		\citenamefont {Frunzio}, \citenamefont {Girvin}, \citenamefont {Jiang} 
		\emph
		{et~al.}}]{ofek2016extending}%
	\BibitemOpen
	\bibfield  {author} {\bibinfo {author} {\bibfnamefont {N.}~\bibnamefont
			{Ofek}}, \bibinfo {author} {\bibfnamefont {A.}~\bibnamefont 
			{Petrenko}},
		\bibinfo {author} {\bibfnamefont {R.}~\bibnamefont {Heeres}}, \bibinfo
		{author} {\bibfnamefont {P.}~\bibnamefont {Reinhold}}, \bibinfo {author}
		{\bibfnamefont {Z.}~\bibnamefont {Leghtas}}, \bibinfo {author} 
		{\bibfnamefont
			{B.}~\bibnamefont {Vlastakis}}, \bibinfo {author} {\bibfnamefont
			{Y.}~\bibnamefont {Liu}}, \bibinfo {author} {\bibfnamefont 
			{L.}~\bibnamefont
			{Frunzio}}, \bibinfo {author} {\bibfnamefont {S.}~\bibnamefont 
			{Girvin}},
		\bibinfo {author} {\bibfnamefont {L.}~\bibnamefont {Jiang}},  \emph
		{et~al.},\ }\href@noop {} {\bibfield  {journal} {\bibinfo  {journal}
			{Nature}\ }\textbf {\bibinfo {volume} {536}},\ \bibinfo {pages} {441}
		(\bibinfo {year} {2016})}\BibitemShut {NoStop}%
	\bibitem [{\citenamefont {Barends}\ \emph {et~al.}(2014)\citenamefont
		{Barends}, \citenamefont {Kelly}, \citenamefont {Megrant}, \citenamefont
		{Veitia}, \citenamefont {Sank}, \citenamefont {Jeffrey}, \citenamefont
		{White}, \citenamefont {Mutus}, \citenamefont {Fowler}, \citenamefont
		{Campbell} \emph {et~al.}}]{barends2014superconducting}%
	\BibitemOpen
	\bibfield  {author} {\bibinfo {author} {\bibfnamefont {R.}~\bibnamefont
			{Barends}}, \bibinfo {author} {\bibfnamefont {J.}~\bibnamefont 
			{Kelly}},
		\bibinfo {author} {\bibfnamefont {A.}~\bibnamefont {Megrant}}, \bibinfo
		{author} {\bibfnamefont {A.}~\bibnamefont {Veitia}}, \bibinfo {author}
		{\bibfnamefont {D.}~\bibnamefont {Sank}}, \bibinfo {author} {\bibfnamefont
			{E.}~\bibnamefont {Jeffrey}}, \bibinfo {author} {\bibfnamefont 
			{T.~C.}\
			\bibnamefont {White}}, \bibinfo {author} {\bibfnamefont 
			{J.}~\bibnamefont
			{Mutus}}, \bibinfo {author} {\bibfnamefont {A.~G.}\ \bibnamefont 
			{Fowler}},
		\bibinfo {author} {\bibfnamefont {B.}~\bibnamefont {Campbell}},  \emph
		{et~al.},\ }\href@noop {} {\bibfield  {journal} {\bibinfo  {journal}
			{Nature}\ }\textbf {\bibinfo {volume} {508}},\ \bibinfo {pages} {500}
		(\bibinfo {year} {2014})}\BibitemShut {NoStop}%
	\bibitem [{\citenamefont {Riste}\ \emph {et~al.}(2013)\citenamefont {Riste},
		\citenamefont {Dukalski}, \citenamefont {Watson}, \citenamefont 
		{De~Lange},
		\citenamefont {Tiggelman}, \citenamefont {Blanter}, \citenamefont 
		{Lehnert},
		\citenamefont {Schouten},\ and\ \citenamefont
		{DiCarlo}}]{riste2013deterministic}%
	\BibitemOpen
	\bibfield  {author} {\bibinfo {author} {\bibfnamefont {D.}~\bibnamefont
			{Riste}}, \bibinfo {author} {\bibfnamefont {M.}~\bibnamefont 
			{Dukalski}},
		\bibinfo {author} {\bibfnamefont {C.}~\bibnamefont {Watson}}, \bibinfo
		{author} {\bibfnamefont {G.}~\bibnamefont {De~Lange}}, \bibinfo {author}
		{\bibfnamefont {M.}~\bibnamefont {Tiggelman}}, \bibinfo {author}
		{\bibfnamefont {Y.~M.}\ \bibnamefont {Blanter}}, \bibinfo {author}
		{\bibfnamefont {K.~W.}\ \bibnamefont {Lehnert}}, \bibinfo {author}
		{\bibfnamefont {R.}~\bibnamefont {Schouten}}, \ and\ \bibinfo {author}
		{\bibfnamefont {L.}~\bibnamefont {DiCarlo}},\ }\href@noop {} {\bibfield
		{journal} {\bibinfo  {journal} {Nature}\ }\textbf {\bibinfo {volume} 
		{502}},\
		\bibinfo {pages} {350} (\bibinfo {year} {2013})}\BibitemShut {NoStop}%
	\bibitem [{\citenamefont {Vijay}\ \emph {et~al.}(2012)\citenamefont {Vijay},
		\citenamefont {Macklin}, \citenamefont {Slichter}, \citenamefont {Weber},
		\citenamefont {Murch}, \citenamefont {Naik}, \citenamefont {Korotkov},\ 
		and\
		\citenamefont {Siddiqi}}]{vijay2012stabilizing}%
	\BibitemOpen
	\bibfield  {author} {\bibinfo {author} {\bibfnamefont {R.}~\bibnamefont
			{Vijay}}, \bibinfo {author} {\bibfnamefont {C.}~\bibnamefont 
			{Macklin}},
		\bibinfo {author} {\bibfnamefont {D.}~\bibnamefont {Slichter}}, \bibinfo
		{author} {\bibfnamefont {S.}~\bibnamefont {Weber}}, \bibinfo {author}
		{\bibfnamefont {K.}~\bibnamefont {Murch}}, \bibinfo {author} 
		{\bibfnamefont
			{R.}~\bibnamefont {Naik}}, \bibinfo {author} {\bibfnamefont {A.~N.}\
			\bibnamefont {Korotkov}}, \ and\ \bibinfo {author} {\bibfnamefont
			{I.}~\bibnamefont {Siddiqi}},\ }\href@noop {} {\bibfield  {journal} 
			{\bibinfo
			{journal} {Nature}\ }\textbf {\bibinfo {volume} {490}},\ \bibinfo 
			{pages}
		{77} (\bibinfo {year} {2012})}\BibitemShut {NoStop}%
	\bibitem [{\citenamefont {Yamamoto}\ \emph {et~al.}(2008)\citenamefont
		{Yamamoto}, \citenamefont {Inomata}, \citenamefont {Watanabe}, 
		\citenamefont
		{Matsuba}, \citenamefont {Miyazaki}, \citenamefont {Oliver}, \citenamefont
		{Nakamura},\ and\ \citenamefont {Tsai}}]{yamamoto2008flux}%
	\BibitemOpen
	\bibfield  {author} {\bibinfo {author} {\bibfnamefont {T.}~\bibnamefont
			{Yamamoto}}, \bibinfo {author} {\bibfnamefont {K.}~\bibnamefont 
			{Inomata}},
		\bibinfo {author} {\bibfnamefont {M.}~\bibnamefont {Watanabe}}, \bibinfo
		{author} {\bibfnamefont {K.}~\bibnamefont {Matsuba}}, \bibinfo {author}
		{\bibfnamefont {T.}~\bibnamefont {Miyazaki}}, \bibinfo {author}
		{\bibfnamefont {W.}~\bibnamefont {Oliver}}, \bibinfo {author} 
		{\bibfnamefont
			{Y.}~\bibnamefont {Nakamura}}, \ and\ \bibinfo {author} {\bibfnamefont
			{J.}~\bibnamefont {Tsai}},\ }\href@noop {} {\bibfield  {journal} 
			{\bibinfo
			{journal} {Applied Physics Letters}\ }\textbf {\bibinfo {volume} 
			{93}},\
		\bibinfo {pages} {042510} (\bibinfo {year} {2008})}\BibitemShut {NoStop}%
	\bibitem [{\citenamefont {Castellanos-Beltran}\ \emph
		{et~al.}(2008)\citenamefont {Castellanos-Beltran}, \citenamefont {Irwin},
		\citenamefont {Hilton}, \citenamefont {Vale},\ and\ \citenamefont
		{Lehnert}}]{castellanos2008amplification}%
	\BibitemOpen
	\bibfield  {author} {\bibinfo {author} {\bibfnamefont {M.}~\bibnamefont
			{Castellanos-Beltran}}, \bibinfo {author} {\bibfnamefont 
			{K.}~\bibnamefont
			{Irwin}}, \bibinfo {author} {\bibfnamefont {G.}~\bibnamefont 
			{Hilton}},
		\bibinfo {author} {\bibfnamefont {L.}~\bibnamefont {Vale}}, \ and\ 
		\bibinfo
		{author} {\bibfnamefont {K.}~\bibnamefont {Lehnert}},\ }\href@noop {}
	{\bibfield  {journal} {\bibinfo  {journal} {Nature Physics}\ }\textbf
		{\bibinfo {volume} {4}},\ \bibinfo {pages} {929} (\bibinfo {year}
		{2008})}\BibitemShut {NoStop}%
	\bibitem [{\citenamefont {Bergeal}\ \emph {et~al.}(2010)\citenamefont
		{Bergeal}, \citenamefont {Schackert}, \citenamefont {Metcalfe}, 
		\citenamefont
		{Vijay}, \citenamefont {Manucharyan}, \citenamefont {Frunzio}, 
		\citenamefont
		{Prober}, \citenamefont {Schoelkopf}, \citenamefont {Girvin},\ and\
		\citenamefont {Devoret}}]{bergeal2010phase}%
	\BibitemOpen
	\bibfield  {author} {\bibinfo {author} {\bibfnamefont {N.}~\bibnamefont
			{Bergeal}}, \bibinfo {author} {\bibfnamefont {F.}~\bibnamefont 
			{Schackert}},
		\bibinfo {author} {\bibfnamefont {M.}~\bibnamefont {Metcalfe}}, \bibinfo
		{author} {\bibfnamefont {R.}~\bibnamefont {Vijay}}, \bibinfo {author}
		{\bibfnamefont {V.}~\bibnamefont {Manucharyan}}, \bibinfo {author}
		{\bibfnamefont {L.}~\bibnamefont {Frunzio}}, \bibinfo {author} 
		{\bibfnamefont
			{D.}~\bibnamefont {Prober}}, \bibinfo {author} {\bibfnamefont
			{R.}~\bibnamefont {Schoelkopf}}, \bibinfo {author} {\bibfnamefont
			{S.}~\bibnamefont {Girvin}}, \ and\ \bibinfo {author} {\bibfnamefont
			{M.}~\bibnamefont {Devoret}},\ }\href@noop {} {\bibfield  {journal} 
			{\bibinfo
			{journal} {Nature}\ }\textbf {\bibinfo {volume} {465}},\ \bibinfo 
			{pages}
		{64} (\bibinfo {year} {2010})}\BibitemShut {NoStop}%
	\bibitem [{\citenamefont {Lecocq}\ \emph {et~al.}(2017)\citenamefont {Lecocq},
		\citenamefont {Ranzani}, \citenamefont {Peterson}, \citenamefont {Cicak},
		\citenamefont {Simmonds}, \citenamefont {Teufel},\ and\ \citenamefont
		{Aumentado}}]{lecocq2017nonreciprocal}%
	\BibitemOpen
	\bibfield  {author} {\bibinfo {author} {\bibfnamefont {F.}~\bibnamefont
			{Lecocq}}, \bibinfo {author} {\bibfnamefont {L.}~\bibnamefont 
			{Ranzani}},
		\bibinfo {author} {\bibfnamefont {G.}~\bibnamefont {Peterson}}, \bibinfo
		{author} {\bibfnamefont {K.}~\bibnamefont {Cicak}}, \bibinfo {author}
		{\bibfnamefont {R.}~\bibnamefont {Simmonds}}, \bibinfo {author}
		{\bibfnamefont {J.}~\bibnamefont {Teufel}}, \ and\ \bibinfo {author}
		{\bibfnamefont {J.}~\bibnamefont {Aumentado}},\ }\href@noop {} {\bibfield
		{journal} {\bibinfo  {journal} {Physical Review Applied}\ }\textbf 
		{\bibinfo
			{volume} {7}},\ \bibinfo {pages} {024028} (\bibinfo {year}
		{2017})}\BibitemShut {NoStop}%
	\bibitem [{\citenamefont {Mutus}\ \emph {et~al.}(2014)\citenamefont {Mutus},
		\citenamefont {White}, \citenamefont {Barends}, \citenamefont {Chen},
		\citenamefont {Chen}, \citenamefont {Chiaro}, \citenamefont {Dunsworth},
		\citenamefont {Jeffrey}, \citenamefont {Kelly}, \citenamefont {Megrant} 
		\emph
		{et~al.}}]{mutus2014strong}%
	\BibitemOpen
	\bibfield  {author} {\bibinfo {author} {\bibfnamefont {J.~Y.}\ \bibnamefont
			{Mutus}}, \bibinfo {author} {\bibfnamefont {T.~C.}\ \bibnamefont 
			{White}},
		\bibinfo {author} {\bibfnamefont {R.}~\bibnamefont {Barends}}, \bibinfo
		{author} {\bibfnamefont {Y.}~\bibnamefont {Chen}}, \bibinfo {author}
		{\bibfnamefont {Z.}~\bibnamefont {Chen}}, \bibinfo {author} {\bibfnamefont
			{B.}~\bibnamefont {Chiaro}}, \bibinfo {author} {\bibfnamefont
			{A.}~\bibnamefont {Dunsworth}}, \bibinfo {author} {\bibfnamefont
			{E.}~\bibnamefont {Jeffrey}}, \bibinfo {author} {\bibfnamefont
			{J.}~\bibnamefont {Kelly}}, \bibinfo {author} {\bibfnamefont
			{A.}~\bibnamefont {Megrant}},  \emph {et~al.},\ }\href@noop {} 
			{\bibfield
		{journal} {\bibinfo  {journal} {Applied Physics Letters}\ }\textbf 
		{\bibinfo
			{volume} {104}},\ \bibinfo {pages} {263513} (\bibinfo {year}
		{2014})}\BibitemShut {NoStop}%
	\bibitem [{\citenamefont {Eichler}\ \emph {et~al.}(2014)\citenamefont
		{Eichler}, \citenamefont {Salathe}, \citenamefont {Mlynek}, \citenamefont
		{Schmidt},\ and\ \citenamefont {Wallraff}}]{eichler2014quantum}%
	\BibitemOpen
	\bibfield  {author} {\bibinfo {author} {\bibfnamefont {C.}~\bibnamefont
			{Eichler}}, \bibinfo {author} {\bibfnamefont {Y.}~\bibnamefont 
			{Salathe}},
		\bibinfo {author} {\bibfnamefont {J.}~\bibnamefont {Mlynek}}, \bibinfo
		{author} {\bibfnamefont {S.}~\bibnamefont {Schmidt}}, \ and\ \bibinfo
		{author} {\bibfnamefont {A.}~\bibnamefont {Wallraff}},\ }\href@noop {}
	{\bibfield  {journal} {\bibinfo  {journal} {Physical review letters}\
		}\textbf {\bibinfo {volume} {113}},\ \bibinfo {pages} {110502} (\bibinfo
		{year} {2014})}\BibitemShut {NoStop}%
	\bibitem [{\citenamefont {Roy}\ \emph {et~al.}(2015)\citenamefont {Roy},
		\citenamefont {Kundu}, \citenamefont {Chand}, \citenamefont {Vadiraj},
		\citenamefont {Ranadive}, \citenamefont {Nehra}, \citenamefont {Patankar},
		\citenamefont {Aumentado}, \citenamefont {Clerk},\ and\ \citenamefont
		{Vijay}}]{roy2015broadband}%
	\BibitemOpen
	\bibfield  {author} {\bibinfo {author} {\bibfnamefont {T.}~\bibnamefont
			{Roy}}, \bibinfo {author} {\bibfnamefont {S.}~\bibnamefont {Kundu}}, 
			\bibinfo
		{author} {\bibfnamefont {M.}~\bibnamefont {Chand}}, \bibinfo {author}
		{\bibfnamefont {A.}~\bibnamefont {Vadiraj}}, \bibinfo {author} 
		{\bibfnamefont
			{A.}~\bibnamefont {Ranadive}}, \bibinfo {author} {\bibfnamefont
			{N.}~\bibnamefont {Nehra}}, \bibinfo {author} {\bibfnamefont {M.~P.}\
			\bibnamefont {Patankar}}, \bibinfo {author} {\bibfnamefont 
			{J.}~\bibnamefont
			{Aumentado}}, \bibinfo {author} {\bibfnamefont {A.}~\bibnamefont 
			{Clerk}}, \
		and\ \bibinfo {author} {\bibfnamefont {R.}~\bibnamefont {Vijay}},\
	}\href@noop {} {\bibfield  {journal} {\bibinfo  {journal} {Applied Physics
				Letters}\ }\textbf {\bibinfo {volume} {107}},\ \bibinfo {pages} 
				{262601}
		(\bibinfo {year} {2015})}\BibitemShut {NoStop}%
	\bibitem [{\citenamefont {Simoen}\ \emph {et~al.}(2015)\citenamefont {Simoen},
		\citenamefont {Chang}, \citenamefont {Krantz}, \citenamefont {Bylander},
		\citenamefont {Wustmann}, \citenamefont {Shumeiko}, \citenamefont 
		{Delsing},\
		and\ \citenamefont {Wilson}}]{simoen2015characterization}%
	\BibitemOpen
	\bibfield  {author} {\bibinfo {author} {\bibfnamefont {M.}~\bibnamefont
			{Simoen}}, \bibinfo {author} {\bibfnamefont {C.}~\bibnamefont 
			{Chang}},
		\bibinfo {author} {\bibfnamefont {P.}~\bibnamefont {Krantz}}, \bibinfo
		{author} {\bibfnamefont {J.}~\bibnamefont {Bylander}}, \bibinfo {author}
		{\bibfnamefont {W.}~\bibnamefont {Wustmann}}, \bibinfo {author}
		{\bibfnamefont {V.}~\bibnamefont {Shumeiko}}, \bibinfo {author}
		{\bibfnamefont {P.}~\bibnamefont {Delsing}}, \ and\ \bibinfo {author}
		{\bibfnamefont {C.}~\bibnamefont {Wilson}},\ }\href@noop {} {\bibfield
		{journal} {\bibinfo  {journal} {Journal of Applied Physics}\ }\textbf
		{\bibinfo {volume} {118}},\ \bibinfo {pages} {154501} (\bibinfo {year}
		{2015})}\BibitemShut {NoStop}%
	\bibitem [{\citenamefont {Chen}\ \emph {et~al.}(2014)\citenamefont {Chen},
		\citenamefont {Neill}, \citenamefont {Roushan}, \citenamefont {Leung},
		\citenamefont {Fang}, \citenamefont {Barends}, \citenamefont {Kelly},
		\citenamefont {Campbell}, \citenamefont {Chen}, \citenamefont {Chiaro} 
		\emph
		{et~al.}}]{chen2014qubit}%
	\BibitemOpen
	\bibfield  {author} {\bibinfo {author} {\bibfnamefont {Y.}~\bibnamefont
			{Chen}}, \bibinfo {author} {\bibfnamefont {C.}~\bibnamefont {Neill}},
		\bibinfo {author} {\bibfnamefont {P.}~\bibnamefont {Roushan}}, \bibinfo
		{author} {\bibfnamefont {N.}~\bibnamefont {Leung}}, \bibinfo {author}
		{\bibfnamefont {M.}~\bibnamefont {Fang}}, \bibinfo {author} {\bibfnamefont
			{R.}~\bibnamefont {Barends}}, \bibinfo {author} {\bibfnamefont
			{J.}~\bibnamefont {Kelly}}, \bibinfo {author} {\bibfnamefont
			{B.}~\bibnamefont {Campbell}}, \bibinfo {author} {\bibfnamefont
			{Z.}~\bibnamefont {Chen}}, \bibinfo {author} {\bibfnamefont 
			{B.}~\bibnamefont
			{Chiaro}},  \emph {et~al.},\ }\href@noop {} {\bibfield  {journal} 
			{\bibinfo
			{journal} {Physical review letters}\ }\textbf {\bibinfo {volume} 
			{113}},\
		\bibinfo {pages} {220502} (\bibinfo {year} {2014})}\BibitemShut {NoStop}%
	\bibitem [{\citenamefont {Takita}\ \emph {et~al.}(2017)\citenamefont {Takita},
		\citenamefont {Cross}, \citenamefont {C{\'o}rcoles}, \citenamefont 
		{Chow},\
		and\ \citenamefont {Gambetta}}]{takita2017experimental}%
	\BibitemOpen
	\bibfield  {author} {\bibinfo {author} {\bibfnamefont {M.}~\bibnamefont
			{Takita}}, \bibinfo {author} {\bibfnamefont {A.~W.}\ \bibnamefont 
			{Cross}},
		\bibinfo {author} {\bibfnamefont {A.}~\bibnamefont {C{\'o}rcoles}}, 
		\bibinfo
		{author} {\bibfnamefont {J.~M.}\ \bibnamefont {Chow}}, \ and\ \bibinfo
		{author} {\bibfnamefont {J.~M.}\ \bibnamefont {Gambetta}},\ }\href@noop {}
	{\bibfield  {journal} {\bibinfo  {journal} {Physical review letters}\
		}\textbf {\bibinfo {volume} {119}},\ \bibinfo {pages} {180501} (\bibinfo
		{year} {2017})}\BibitemShut {NoStop}%
	\bibitem [{\citenamefont {Macklin}\ \emph {et~al.}(2015)\citenamefont
		{Macklin}, \citenamefont {O’Brien}, \citenamefont {Hover}, \citenamefont
		{Schwartz}, \citenamefont {Bolkhovsky}, \citenamefont {Zhang}, 
		\citenamefont
		{Oliver},\ and\ \citenamefont {Siddiqi}}]{macklin2015near}%
	\BibitemOpen
	\bibfield  {author} {\bibinfo {author} {\bibfnamefont {C.}~\bibnamefont
			{Macklin}}, \bibinfo {author} {\bibfnamefont {K.}~\bibnamefont 
			{O’Brien}},
		\bibinfo {author} {\bibfnamefont {D.}~\bibnamefont {Hover}}, \bibinfo
		{author} {\bibfnamefont {M.}~\bibnamefont {Schwartz}}, \bibinfo {author}
		{\bibfnamefont {V.}~\bibnamefont {Bolkhovsky}}, \bibinfo {author}
		{\bibfnamefont {X.}~\bibnamefont {Zhang}}, \bibinfo {author} 
		{\bibfnamefont
			{W.}~\bibnamefont {Oliver}}, \ and\ \bibinfo {author} {\bibfnamefont
			{I.}~\bibnamefont {Siddiqi}},\ }\href@noop {} {\bibfield  {journal} 
			{\bibinfo
			{journal} {Science}\ }\textbf {\bibinfo {volume} {350}},\ \bibinfo 
			{pages}
		{307} (\bibinfo {year} {2015})}\BibitemShut {NoStop}%
	\bibitem [{\citenamefont {O’Brien}\ \emph {et~al.}(2014)\citenamefont
		{O’Brien}, \citenamefont {Macklin}, \citenamefont {Siddiqi},\ and\
		\citenamefont {Zhang}}]{o2014resonant}%
	\BibitemOpen
	\bibfield  {author} {\bibinfo {author} {\bibfnamefont {K.}~\bibnamefont
			{O’Brien}}, \bibinfo {author} {\bibfnamefont {C.}~\bibnamefont 
			{Macklin}},
		\bibinfo {author} {\bibfnamefont {I.}~\bibnamefont {Siddiqi}}, \ and\
		\bibinfo {author} {\bibfnamefont {X.}~\bibnamefont {Zhang}},\ }\href@noop 
		{}
	{\bibfield  {journal} {\bibinfo  {journal} {Physical Review Letters}\
		}\textbf {\bibinfo {volume} {113}},\ \bibinfo {pages} {157001} (\bibinfo
		{year} {2014})}\BibitemShut {NoStop}%
	\bibitem [{\citenamefont {Bell}\ and\ \citenamefont
		{Samolov}(2015)}]{bell2015traveling}%
	\BibitemOpen
	\bibfield  {author} {\bibinfo {author} {\bibfnamefont {M.}~\bibnamefont
			{Bell}}\ and\ \bibinfo {author} {\bibfnamefont {A.}~\bibnamefont 
			{Samolov}},\
	}\href@noop {} {\bibfield  {journal} {\bibinfo  {journal} {Physical Review
				Applied}\ }\textbf {\bibinfo {volume} {4}},\ \bibinfo {pages} 
				{024014}
		(\bibinfo {year} {2015})}\BibitemShut {NoStop}%
	\bibitem [{\citenamefont {White}\ \emph {et~al.}(2015)\citenamefont {White},
		\citenamefont {Mutus}, \citenamefont {Hoi}, \citenamefont {Barends},
		\citenamefont {Campbell}, \citenamefont {Chen}, \citenamefont {Chen},
		\citenamefont {Chiaro}, \citenamefont {Dunsworth}, \citenamefont {Jeffrey}
		\emph {et~al.}}]{white2015traveling}%
	\BibitemOpen
	\bibfield  {author} {\bibinfo {author} {\bibfnamefont {T.}~\bibnamefont
			{White}}, \bibinfo {author} {\bibfnamefont {J.}~\bibnamefont {Mutus}},
		\bibinfo {author} {\bibfnamefont {I.-C.}\ \bibnamefont {Hoi}}, \bibinfo
		{author} {\bibfnamefont {R.}~\bibnamefont {Barends}}, \bibinfo {author}
		{\bibfnamefont {B.}~\bibnamefont {Campbell}}, \bibinfo {author}
		{\bibfnamefont {Y.}~\bibnamefont {Chen}}, \bibinfo {author} {\bibfnamefont
			{Z.}~\bibnamefont {Chen}}, \bibinfo {author} {\bibfnamefont 
			{B.}~\bibnamefont
			{Chiaro}}, \bibinfo {author} {\bibfnamefont {A.}~\bibnamefont 
			{Dunsworth}},
		\bibinfo {author} {\bibfnamefont {E.}~\bibnamefont {Jeffrey}},  \emph
		{et~al.},\ }\href@noop {} {\bibfield  {journal} {\bibinfo  {journal} 
		{Applied
				Physics Letters}\ }\textbf {\bibinfo {volume} {106}},\ \bibinfo 
				{pages}
		{242601} (\bibinfo {year} {2015})}\BibitemShut {NoStop}%
	\bibitem [{\citenamefont {Zorin}(2016)}]{zorin2016josephson}%
	\BibitemOpen
	\bibfield  {author} {\bibinfo {author} {\bibfnamefont {A.}~\bibnamefont
			{Zorin}},\ }\href@noop {} {\bibfield  {journal} {\bibinfo  {journal}
			{Physical Review Applied}\ }\textbf {\bibinfo {volume} {6}},\ \bibinfo
		{pages} {034006} (\bibinfo {year} {2016})}\BibitemShut {NoStop}%
	\bibitem [{\citenamefont {Zorin}\ \emph {et~al.}(2017)\citenamefont {Zorin},
		\citenamefont {Khabipov}, \citenamefont {Dietel},\ and\ \citenamefont
		{Dolata}}]{zorin2017traveling}%
	\BibitemOpen
	\bibfield  {author} {\bibinfo {author} {\bibfnamefont {A.}~\bibnamefont
			{Zorin}}, \bibinfo {author} {\bibfnamefont {M.}~\bibnamefont 
			{Khabipov}},
		\bibinfo {author} {\bibfnamefont {J.}~\bibnamefont {Dietel}}, \ and\ 
		\bibinfo
		{author} {\bibfnamefont {R.}~\bibnamefont {Dolata}},\ }in\ \href@noop {}
	{\emph {\bibinfo {booktitle} {Superconductive Electronics Conference (ISEC),
				2017 16th International}}}\ (\bibinfo {organization} {IEEE},\ 
				\bibinfo {year}
	{2017})\ pp.\ \bibinfo {pages} {1--3}\BibitemShut {NoStop}%
	\bibitem [{\citenamefont {Eom}\ \emph {et~al.}(2012)\citenamefont {Eom},
		\citenamefont {Day}, \citenamefont {Leduc},\ and\ \citenamefont
		{Zmuidzinas}}]{eom2012wideband}%
	\BibitemOpen
	\bibfield  {author} {\bibinfo {author} {\bibfnamefont {B.~H.}\ \bibnamefont
			{Eom}}, \bibinfo {author} {\bibfnamefont {P.~K.}\ \bibnamefont {Day}},
		\bibinfo {author} {\bibfnamefont {H.~G.}\ \bibnamefont {Leduc}}, \ and\
		\bibinfo {author} {\bibfnamefont {J.}~\bibnamefont {Zmuidzinas}},\
	}\href@noop {} {\bibfield  {journal} {\bibinfo  {journal} {Nature Physics}\
		}\textbf {\bibinfo {volume} {8}},\ \bibinfo {pages} {623} (\bibinfo {year}
		{2012})}\BibitemShut {NoStop}%
	\bibitem [{\citenamefont {Vissers}\ \emph {et~al.}(2016)\citenamefont
		{Vissers}, \citenamefont {Erickson}, \citenamefont {Ku}, \citenamefont
		{Vale}, \citenamefont {Wu}, \citenamefont {Hilton},\ and\ \citenamefont
		{Pappas}}]{vissers2016low}%
	\BibitemOpen
	\bibfield  {author} {\bibinfo {author} {\bibfnamefont {M.~R.}\ \bibnamefont
			{Vissers}}, \bibinfo {author} {\bibfnamefont {R.~P.}\ \bibnamefont
			{Erickson}}, \bibinfo {author} {\bibfnamefont {H.-S.}\ \bibnamefont 
			{Ku}},
		\bibinfo {author} {\bibfnamefont {L.}~\bibnamefont {Vale}}, \bibinfo 
		{author}
		{\bibfnamefont {X.}~\bibnamefont {Wu}}, \bibinfo {author} {\bibfnamefont
			{G.}~\bibnamefont {Hilton}}, \ and\ \bibinfo {author} {\bibfnamefont 
			{D.~P.}\
			\bibnamefont {Pappas}},\ }\href@noop {} {\bibfield  {journal} 
			{\bibinfo
			{journal} {Applied physics letters}\ }\textbf {\bibinfo {volume} 
			{108}},\
		\bibinfo {pages} {012601} (\bibinfo {year} {2016})}\BibitemShut {NoStop}%
	\bibitem [{\citenamefont {Erickson}\ and\ \citenamefont
		{Pappas}(2017)}]{erickson2017theory}%
	\BibitemOpen
	\bibfield  {author} {\bibinfo {author} {\bibfnamefont {R.~P.}\ \bibnamefont
			{Erickson}}\ and\ \bibinfo {author} {\bibfnamefont {D.~P.}\ 
			\bibnamefont
			{Pappas}},\ }\href@noop {} {\bibfield  {journal} {\bibinfo  {journal}
			{Physical Review B}\ }\textbf {\bibinfo {volume} {95}},\ \bibinfo 
			{pages}
		{104506} (\bibinfo {year} {2017})}\BibitemShut {NoStop}%
	\bibitem [{\citenamefont {Chaudhuri}\ \emph {et~al.}(2017)\citenamefont
		{Chaudhuri}, \citenamefont {Li}, \citenamefont {Irwin}, \citenamefont
		{Bockstiegel}, \citenamefont {Hubmayr}, \citenamefont {Ullom}, 
		\citenamefont
		{Vissers},\ and\ \citenamefont {Gao}}]{chaudhuri2017broadband}%
	\BibitemOpen
	\bibfield  {author} {\bibinfo {author} {\bibfnamefont {S.}~\bibnamefont
			{Chaudhuri}}, \bibinfo {author} {\bibfnamefont {D.}~\bibnamefont 
			{Li}},
		\bibinfo {author} {\bibfnamefont {K.}~\bibnamefont {Irwin}}, \bibinfo
		{author} {\bibfnamefont {C.}~\bibnamefont {Bockstiegel}}, \bibinfo 
		{author}
		{\bibfnamefont {J.}~\bibnamefont {Hubmayr}}, \bibinfo {author} 
		{\bibfnamefont
			{J.}~\bibnamefont {Ullom}}, \bibinfo {author} {\bibfnamefont
			{M.}~\bibnamefont {Vissers}}, \ and\ \bibinfo {author} {\bibfnamefont
			{J.}~\bibnamefont {Gao}},\ }\href@noop {} {\bibfield  {journal} 
			{\bibinfo
			{journal} {Applied Physics Letters}\ }\textbf {\bibinfo {volume} 
			{110}},\
		\bibinfo {pages} {152601} (\bibinfo {year} {2017})}\BibitemShut {NoStop}%
	\bibitem [{\citenamefont {Hunter}(2001)}]{hunter2001theory}%
	\BibitemOpen
	\bibfield  {author} {\bibinfo {author} {\bibfnamefont {I.}~\bibnamefont
			{Hunter}},\ }\href@noop {} {\emph {\bibinfo {title} {Theory and 
			design of
				microwave filters}}},\ \bibinfo {number} {48}\ (\bibinfo  
				{publisher} {Iet},\
	\bibinfo {year} {2001})\BibitemShut {NoStop}%
	\bibitem [{\citenamefont {Ranzani}\ \emph {et~al.}(2013)\citenamefont
		{Ranzani}, \citenamefont {Spietz}, \citenamefont {Popovic},\ and\
		\citenamefont {Aumentado}}]{ranzani2013two}%
	\BibitemOpen
	\bibfield  {author} {\bibinfo {author} {\bibfnamefont {L.}~\bibnamefont
			{Ranzani}}, \bibinfo {author} {\bibfnamefont {L.}~\bibnamefont 
			{Spietz}},
		\bibinfo {author} {\bibfnamefont {Z.}~\bibnamefont {Popovic}}, \ and\
		\bibinfo {author} {\bibfnamefont {J.}~\bibnamefont {Aumentado}},\ 
		}\href@noop
	{} {\bibfield  {journal} {\bibinfo  {journal} {Review of Scientific
				Instruments}\ }\textbf {\bibinfo {volume} {84}},\ \bibinfo 
				{pages} {034704}
		(\bibinfo {year} {2013})}\BibitemShut {NoStop}%
	\bibitem [{\citenamefont {Jeffrey}\ \emph {et~al.}(2014)\citenamefont
		{Jeffrey}, \citenamefont {Sank}, \citenamefont {Mutus}, \citenamefont
		{White}, \citenamefont {Kelly}, \citenamefont {Barends}, \citenamefont
		{Chen}, \citenamefont {Chen}, \citenamefont {Chiaro}, \citenamefont
		{Dunsworth} \emph {et~al.}}]{jeffrey2014fast}%
	\BibitemOpen
	\bibfield  {author} {\bibinfo {author} {\bibfnamefont {E.}~\bibnamefont
			{Jeffrey}}, \bibinfo {author} {\bibfnamefont {D.}~\bibnamefont 
			{Sank}},
		\bibinfo {author} {\bibfnamefont {J.}~\bibnamefont {Mutus}}, \bibinfo
		{author} {\bibfnamefont {T.}~\bibnamefont {White}}, \bibinfo {author}
		{\bibfnamefont {J.}~\bibnamefont {Kelly}}, \bibinfo {author} 
		{\bibfnamefont
			{R.}~\bibnamefont {Barends}}, \bibinfo {author} {\bibfnamefont
			{Y.}~\bibnamefont {Chen}}, \bibinfo {author} {\bibfnamefont 
			{Z.}~\bibnamefont
			{Chen}}, \bibinfo {author} {\bibfnamefont {B.}~\bibnamefont {Chiaro}},
		\bibinfo {author} {\bibfnamefont {A.}~\bibnamefont {Dunsworth}},  \emph
		{et~al.},\ }\href@noop {} {\bibfield  {journal} {\bibinfo  {journal}
			{Physical review letters}\ }\textbf {\bibinfo {volume} {112}},\ 
			\bibinfo
		{pages} {190504} (\bibinfo {year} {2014})}\BibitemShut {NoStop}%
	\bibitem [{\citenamefont {Schreier}\ \emph {et~al.}(2008)\citenamefont
		{Schreier}, \citenamefont {Houck}, \citenamefont {Koch}, \citenamefont
		{Schuster}, \citenamefont {Johnson}, \citenamefont {Chow}, \citenamefont
		{Gambetta}, \citenamefont {Majer}, \citenamefont {Frunzio}, \citenamefont
		{Devoret}, \citenamefont {Girvin},\ and\ \citenamefont
		{Schoelkopf}}]{schreier2008chargenoise}%
	\BibitemOpen
	\bibfield  {author} {\bibinfo {author} {\bibfnamefont {J.~A.}\ \bibnamefont
			{Schreier}}, \bibinfo {author} {\bibfnamefont {A.~A.}\ \bibnamefont 
			{Houck}},
		\bibinfo {author} {\bibfnamefont {J.}~\bibnamefont {Koch}}, \bibinfo 
		{author}
		{\bibfnamefont {D.~I.}\ \bibnamefont {Schuster}}, \bibinfo {author}
		{\bibfnamefont {B.~R.}\ \bibnamefont {Johnson}}, \bibinfo {author}
		{\bibfnamefont {J.~M.}\ \bibnamefont {Chow}}, \bibinfo {author}
		{\bibfnamefont {J.~M.}\ \bibnamefont {Gambetta}}, \bibinfo {author}
		{\bibfnamefont {J.}~\bibnamefont {Majer}}, \bibinfo {author} 
		{\bibfnamefont
			{L.}~\bibnamefont {Frunzio}}, \bibinfo {author} {\bibfnamefont 
			{M.~H.}\
			\bibnamefont {Devoret}}, \bibinfo {author} {\bibfnamefont {S.~M.}\
			\bibnamefont {Girvin}}, \ and\ \bibinfo {author} {\bibfnamefont 
			{R.~J.}\
			\bibnamefont {Schoelkopf}},\ }\href {\doibase 
			10.1103/PhysRevB.77.180502}
	{\bibfield  {journal} {\bibinfo  {journal} {Phys. Rev. B}\ }\textbf {\bibinfo
			{volume} {77}},\ \bibinfo {pages} {180502} (\bibinfo {year}
		{2008})}\BibitemShut {NoStop}%
	\bibitem [{\citenamefont {Ristè}\ \emph {et~al.}(2013)\citenamefont {Ristè},
		\citenamefont {Bultink}, \citenamefont {Tiggelman}, \citenamefont 
		{Schouten},
		\citenamefont {Lehnert},\ and\ \citenamefont
		{DiCarlo}}]{riste2013fluctuations}%
	\BibitemOpen
	\bibfield  {author} {\bibinfo {author} {\bibfnamefont {D.}~\bibnamefont
			{Ristè}}, \bibinfo {author} {\bibfnamefont {C.~C.}\ \bibnamefont 
			{Bultink}},
		\bibinfo {author} {\bibfnamefont {M.~J.}\ \bibnamefont {Tiggelman}}, 
		\bibinfo
		{author} {\bibfnamefont {R.~N.}\ \bibnamefont {Schouten}}, \bibinfo 
		{author}
		{\bibfnamefont {K.~W.}\ \bibnamefont {Lehnert}}, \ and\ \bibinfo {author}
		{\bibfnamefont {L.}~\bibnamefont {DiCarlo}},\ }\href
	{http://dx.doi.org/10.1038/ncomms2936} {\bibfield  {journal} {\bibinfo
			{journal} {Nature Communications}\ }\textbf {\bibinfo {volume} {4}},\
		\bibinfo {pages} {1913} (\bibinfo {year} {2013})}\BibitemShut {NoStop}%
	\bibitem [{\citenamefont {Ryan}\ \emph {et~al.}(2015)\citenamefont {Ryan},
		\citenamefont {Johnson}, \citenamefont {Gambetta}, \citenamefont {Chow},
		\citenamefont {da~Silva}, \citenamefont {Dial},\ and\ \citenamefont
		{Ohki}}]{ryan2015tomo}%
	\BibitemOpen
	\bibfield  {author} {\bibinfo {author} {\bibfnamefont {C.~A.}\ \bibnamefont
			{Ryan}}, \bibinfo {author} {\bibfnamefont {B.~R.}\ \bibnamefont 
			{Johnson}},
		\bibinfo {author} {\bibfnamefont {J.~M.}\ \bibnamefont {Gambetta}}, 
		\bibinfo
		{author} {\bibfnamefont {J.~M.}\ \bibnamefont {Chow}}, \bibinfo {author}
		{\bibfnamefont {M.~P.}\ \bibnamefont {da~Silva}}, \bibinfo {author}
		{\bibfnamefont {O.~E.}\ \bibnamefont {Dial}}, \ and\ \bibinfo {author}
		{\bibfnamefont {T.~A.}\ \bibnamefont {Ohki}},\ }\href {\doibase
		10.1103/PhysRevA.91.022118} {\bibfield  {journal} {\bibinfo  {journal} 
		{Phys.
				Rev. A}\ }\textbf {\bibinfo {volume} {91}},\ \bibinfo {pages} 
				{022118}
		(\bibinfo {year} {2015})}\BibitemShut {NoStop}%
	\bibitem [{\citenamefont {Heinsoo}\ \emph {et~al.}(2018)\citenamefont
		{Heinsoo}, \citenamefont {Andersen}, \citenamefont {Remm}, \citenamefont
		{Krinner}, \citenamefont {Walter}, \citenamefont {Salath\'e}, 
		\citenamefont
		{Gasparinetti}, \citenamefont {Besse}, \citenamefont
		{Poto\ifmmode~\check{c}\else \v{c}\fi{}nik}, \citenamefont {Wallraff},\ 
		and\
		\citenamefont {Eichler}}]{heinsoo2018rapid}%
	\BibitemOpen
	\bibfield  {author} {\bibinfo {author} {\bibfnamefont {J.}~\bibnamefont
			{Heinsoo}}, \bibinfo {author} {\bibfnamefont {C.~K.}\ \bibnamefont
			{Andersen}}, \bibinfo {author} {\bibfnamefont {A.}~\bibnamefont 
			{Remm}},
		\bibinfo {author} {\bibfnamefont {S.}~\bibnamefont {Krinner}}, \bibinfo
		{author} {\bibfnamefont {T.}~\bibnamefont {Walter}}, \bibinfo {author}
		{\bibfnamefont {Y.}~\bibnamefont {Salath\'e}}, \bibinfo {author}
		{\bibfnamefont {S.}~\bibnamefont {Gasparinetti}}, \bibinfo {author}
		{\bibfnamefont {J.-C.}\ \bibnamefont {Besse}}, \bibinfo {author}
		{\bibfnamefont {A.}~\bibnamefont {Poto\ifmmode~\check{c}\else
				\v{c}\fi{}nik}}, \bibinfo {author} {\bibfnamefont 
				{A.}~\bibnamefont
			{Wallraff}}, \ and\ \bibinfo {author} {\bibfnamefont {C.}~\bibnamefont
			{Eichler}},\ }\href {\doibase 10.1103/PhysRevApplied.10.034040} 
			{\bibfield
		{journal} {\bibinfo  {journal} {Phys. Rev. Applied}\ }\textbf {\bibinfo
			{volume} {10}},\ \bibinfo {pages} {034040} (\bibinfo {year}
		{2018})}\BibitemShut {NoStop}%
	\bibitem [{\citenamefont {Gu}\ \emph {et~al.}(2017)\citenamefont {Gu},
		\citenamefont {Kockum}, \citenamefont {Miranowicz}, \citenamefont {Liu},\
		and\ \citenamefont {Nori}}]{gu2017microwave}%
	\BibitemOpen
	\bibfield  {author} {\bibinfo {author} {\bibfnamefont {X.}~\bibnamefont
			{Gu}}, \bibinfo {author} {\bibfnamefont {A.~F.}\ \bibnamefont 
			{Kockum}},
		\bibinfo {author} {\bibfnamefont {A.}~\bibnamefont {Miranowicz}}, \bibinfo
		{author} {\bibfnamefont {Y.-x.}\ \bibnamefont {Liu}}, \ and\ \bibinfo
		{author} {\bibfnamefont {F.}~\bibnamefont {Nori}},\ }\href@noop {} 
		{\bibfield
		{journal} {\bibinfo  {journal} {Physics Reports}\ } (\bibinfo {year}
		{2017})}\BibitemShut {NoStop}%
	\bibitem [{\citenamefont {Adamyan}\ \emph {et~al.}(2016)\citenamefont
		{Adamyan}, \citenamefont {de~Graaf}, \citenamefont {Kubatkin},\ and\
		\citenamefont {Danilov}}]{adamyan2016superconducting}%
	\BibitemOpen
	\bibfield  {author} {\bibinfo {author} {\bibfnamefont {A.}~\bibnamefont
			{Adamyan}}, \bibinfo {author} {\bibfnamefont {S.}~\bibnamefont 
			{de~Graaf}},
		\bibinfo {author} {\bibfnamefont {S.}~\bibnamefont {Kubatkin}}, \ and\
		\bibinfo {author} {\bibfnamefont {A.}~\bibnamefont {Danilov}},\ 
		}\href@noop
	{} {\bibfield  {journal} {\bibinfo  {journal} {Journal of Applied Physics}\
		}\textbf {\bibinfo {volume} {119}},\ \bibinfo {pages} {083901} (\bibinfo
		{year} {2016})}\BibitemShut {NoStop}%
\end{thebibliography}
%
\end{document}